%% file: sample63.tex
\documentclass[twocolumn]{aastex63}

\submitjournal{ApJ}

\usepackage{xspace}

\newcommand{\name}{SN~2019yvq\xspace}
\newcommand{\sne}{SNe Ia\xspace}
\newcommand{\Ha}{\hbox{H$\alpha$}\xspace}
\newcommand{\sn}{SN Ia\xspace}
\newcommand{\CaII}{[\ion{Ca}{2}]\xspace}
\newcommand{\CaIIlong}{[\ion{Ca}{2}]$\lambda7300~\rm\AA$\xspace}
\newcommand{\CaIINIR}{\ion{Ca}{2}\xspace}
\newcommand{\FeII}{[\ion{Fe}{2}]\xspace}
\newcommand{\FeIII}{[\ion{Fe}{3}]\xspace}
\newcommand{\SII}{[\ion{S}{2}]\xspace}
\newcommand{\SiII}{\ion{Si}{2}\xspace}
\newcommand{\CoII}{[\ion{Co}{2}]\xspace}
\newcommand{\CoIII}{[\ion{Co}{3}]\xspace}
\newcommand{\NiII}{[\ion{Ni}{2}]\xspace}
\newcommand{\NiIII}{[\ion{Ni}{3}]\xspace}
\newcommand{\Pa}{Pa$\alpha$\xspace}
\newcommand{\Pb}{Pa$\beta$\xspace}
\newcommand{\kms}{\hbox{km~s$^{-1}$}\xspace}
\newcommand{\um}{\hbox{$\mu\rm m$}\xspace}
\newcommand{\fsNi}{$^{56}$Ni\xspace}
\newcommand{\HeINIRlong}{\ion{He}{1}$\lambda1.083~\um$\xspace}

\shorttitle{SN~2019yvq}
\shortauthors{Tucker et al.}
\graphicspath{{./}{figures/}}

\begin{document}

\title{SN~2019yvq Does Not Conform to SN Ia Explosion Models}

\correspondingauthor{Michael Tucker}
\email{tuckerma95@gmail.com}

\author[0000-0002-2471-8442]{M. A. Tucker}
\altaffiliation{DOE CSGF Fellow}
\affiliation{Institute for Astronomy, 
University of Hawai'i at Manoa, 
2680 Woodlawn Dr. 
Hawai'i, HI 96822, USA}

\author{C. Ashall}
\affiliation{Institute for Astronomy, 
University of Hawai'i at Manoa, 
2680 Woodlawn Dr. 
Hawai'i, HI 96822, USA}

\author{B. J. Shappee}
\affiliation{Institute for Astronomy, 
University of Hawai'i at Manoa, 
2680 Woodlawn Dr. 
Hawai'i, HI 96822, USA}

\author{P. J. Vallely}
\affiliation{Department of Astronomy, 
The Ohio State University, 
140 West 18th Avenue, Columbus, OH 43210, USA
}

\author{C.~S.~Kochanek}
\affiliation{Department of Astronomy, 
The Ohio State University, 
140 West 18th Avenue, Columbus, OH 43210, USA
}
\affiliation{Center for Cosmology and AstroParticle Physics, The Ohio State University, 191 W. Woodruff Ave., Columbus, OH 43210, USA}

\author[0000-0003-1059-9603]{M. E. Huber}
\affiliation{Institute for Astronomy, 
University of Hawai'i at Manoa, 
2680 Woodlawn Dr. 
Hawai'i, HI 96822, USA}

\author[0000-0002-5259-2314]{G. S. Anand}
\affiliation{Institute for Astronomy, 
University of Hawai'i at Manoa, 
2680 Woodlawn Dr. 
Hawai'i, HI 96822, USA}

\author[0000-0002-2021-1863]{J. V. Keane}
\affiliation{Institute for Astronomy, 
University of Hawai'i at Manoa, 
2680 Woodlawn Dr. 
Hawai'i, HI 96822, USA}

\author{E. Y. Hsiao}
\affiliation{Department of Physics, 
Florida State University, 
Tallahassee, FL 32306, USA}

\author[0000-0001-9206-3460]{T.~W.-S.~Holoien}
\altaffiliation{Carnegie Fellow}
\affiliation{The Observatories of the Carnegie Institution for Science, 813 Santa Barbara St., Pasadena, CA 91101, USA}



\begin{abstract}

We present new photometric and spectroscopic observations of SN~2019yvq, a Type Ia supernova (SN Ia) exhibiting several peculiar properties including an excess of UV/optical flux within days of explosion, a high \ion{Si}{2} velocity, and a low peak luminosity. Photometry near the time of first light places new constraints on the rapid rise of the UV/optical flux excess. A near-infrared spectrum at $+173$~days after maximum light places strict limits on the presence of H or He emission, effectively excluding the presence of a nearby non-degenerate star at the time of explosion. New optical spectra, acquired at +128 and +150 days after maximum light, confirm the presence of \CaIIlong and persistent \CaIINIR NIR triplet emission as SN~2019yvq transitions into the nebular phase. The lack of [\ion{O}{1}]$\lambda6300~\rm\AA$ emission disfavors the violent merger of two C/O white dwarfs (WDs) but the merger of a C/O WD with a He WD cannot be excluded. We compare our findings with several models in the literature postulated to explain the early flux excess including double-detonation explosions, $^{56}$Ni mixing into the outer ejecta during ignition, and interaction with H- and He-deficient circumstellar material. Each model may be able to explain both the early flux excess and the nebular \CaII emission, but none of the models can reconcile the high photospheric velocities with the low peak luminosity without introducing new discrepancies. 

\end{abstract}



\section{Introduction} \label{sec:intro}
Type Ia supernovae (SNe Ia) are crucial probes of cosmological parameters \citep[e.g., ][]{riess1998, perlmutter1999} and produce the majority of iron-group elements in the universe \citep[e.g., ][]{iwamoto1999}. They originate from a white dwarf (WD) star \citep{hoyle1960} but the details of how the WD explodes are widely debated (see \citealp{jha2019} and \citealp{maoz2014} for recent reviews). There are two main progenitor theories: the single-degenerate (SD) and double-degenerate (DD) scenarios, depending on whether the companion is a non-degenerate star or a second WD.

The SD channel requires a non-degenerate star to deposit mass onto the surface of the WD \citep{hoyle1960}. The WD gains mass until it destabilises and explodes, although the destabilization mechanism is still unclear. The presence of a nearby non-degenerate star at the time of explosion naturally leads to several observational signatures \citep[e.g., ][]{wheeler1975}. The fast-moving ejecta will impact the star causing shock emission and irregularities in the rising light curve \citep{kasen2010}, strip/ablate material from the stellar surface \citep[e.g., ][]{marietta2000, boehner2017}, and interact with material carried by the stellar winds to produce radio emission \citep{chevalier1982a, chevalier1982b, panagia2006}. Over the past decade, searches for these signatures have mostly returned non-detections such as a lack of bumps in the rising light curves (e.g., \citealp{bianco2011, fausnaugh2019}), limits on nebular \Ha emission from stripped companion material (e.g., \citealp{leonard2007, shappee2013, tucker2020}), and strict constraints on prompt radio (e.g., \citealp{chomiuk2012,chomiuk2016}) and X-ray (e.g., \citealp{margutti2012, margutti2014}) emission. Although the SD scenario has difficulty explaining normal \sne, it can readily account for some peculiar \sne such as those exhibiting interaction with dense circumstellar material \citep[SNe Ia-CSM, ][]{silverman2013}. 

In the DD channel, a second WD destabilizes the more massive WD and induces the explosion. The lack of a non-degenerate star removes many of the predicted observational signatures of the SD scenario, but confirming binary WDs is exceptionally difficult \citep[e.g., ][]{rebassa2019}. There are subtler predictions for DD systems, such as double-peaked emission lines of radioactive decay products \citep{dong2015, vallely2019} and high continuum polarization \citep{Bulla2016}. However, these predictions also depend on the explosion mechanism, a topic related to, but distinct from, the progenitor system.

The explosion mechanism refers to the process of actually destabilizing and igniting the WD. Two common models are the double-detonation \citep{livne1990} and delayed-detonation \citep{khokhlov1993, hoeflich2017} theories. Double detonation refers to the ignition of a surface shell of He which drives a shock wave inwards to detonate the C/O core \citep{livne1990, livne1991}. The near-surface He layer is  acquired from either a companion He star (i.e., the SD channel, \citealp{bildsten2007}) or from a lower-mass WD (i.e., the DD channel, \citealp{fink2007}). 

The delayed-detonation model can also be applied to either progenitor system, requiring only that the primary WD approaches the Chandrasekhar mass ($M_{Ch}\sim 1.4~M_\odot$) via accretion and the explosion is triggered by compressional heating at the center of the WD \citep{hoeflich1996}. There are other explosion mechanisms in the literature, such as gravitationally-confined detonations (GCDs; \citealp{plewa2004}) or pulsational-delayed detonations (PDD; e.g., \citealp{dessart2014}). However, these models are less frequently invoked and we only briefly discuss them in relation to \name. Observational signatures of delayed- and double-detonation scenarios are usually subtle and are best probed by the rising light curve \citep[e.g., ][]{jiang2017, stritzinger2018, jiang2018, polin2019a, bulla2020} or with the structure of the Fe and electron-capture emission lines in nebular-phase spectra \citep[e.g., ][]{botyanszki2017,mazzali2018,polin2019, wilk2020}. 

Finally, two explosion mechanisms unique to the DD channel are violent mergers \citep[e.g., ][]{pakmor2010} and the direct collision of two WDs \citep[e.g., ][]{rosswog2009}. The direct collision can be driven by orbital perturbations of a third \citep[e.g.,][]{thompson2011, katz2012, shappee2013c, antognini2014} or fourth \citep{pejcha2013, fang2018} body. Additional bodies can similarly enhance the violent merger rate but multi-body systems are not necessary. Population synthesis studies suggest violent mergers can account for the majority of \sne \citep[e.g., ][]{ruiter2009} and the Milky Way WD merger rate is consistent with the observed \sne rate \citep[e.g., ][]{maoz2018}. Conversely, the rate of direct collisions is likely too low to explain the normal \sn rate \citep[e.g., ][]{liu2016, antognini2016, toonen2018, hamers2018} but this scenario is considered a viable channel for producing a diverse set of \sn-like transients \citep[e.g., ][]{rosswog2009, vanrossum2016}.

\name, discovered by \citet{itagaki2019}, exhibits peculiar photometric and spectroscopic signatures for a normal \sn. \citet{miller2020} provided an in-depth analysis of the early photometric and spectroscopic evolution, revealing peculiarities such as an excess of UV/optical flux before maximum light, high-velocity \ion{Si}{2} absorption lines, and a low peak luminosity. They discuss several potential progenitor systems and explosion mechanisms but could not reach a definitive interpretation. \citet{miller2020} state that these ambiguities could potentially be resolved once the innermost regions of the ejecta become visible \citep[e.g., ][]{maeda2010,diamond2015}.

\citet{siebert2020} presented optical spectra at $\sim 153$~days after maximum as \name transitioned into the nebular phase. In these late-phase spectra, \name exhibits many spectral features common to \sne such as broad emission lines of \fsNi decay products (i.e., Co and Fe). However, it also exhibited prominent \CaIIlong and \CaIINIR NIR triplet emission which is atypical for \sne. This led to the conclusion that \name stemmed from a double-detonation explosion, with the caveat that the best-fit model has difficulty reproducing the early light curve.

In this paper we provide new data for \name and reassess the viability of various progenitor+explosion theories. \S\ref{sec:methods} outlines our data acquisition and reduction procedures followed by analyses of the photometry and spectra in \S\ref{sec:phot} and \S\ref{sec:spec}, respectively. \S\ref{sec:discuss} discusses various progenitor and explosions scenarios. Finally, in \S\ref{sec:conclusion}, we summarize our results. Throughout this work, we adopt the same host-galaxy parameters as \citet{miller2020}: $z = 0.00888$, $H_0=73~\rm{km/s/Mpc}$, $\mu= 33.14\pm0.11~\rm{mag}$, $D = 42.5\pm1.5~\rm{Mpc}$, and $E(B-V)_{\rm{tot}} = 0.05~\rm{mag}$ (Milky Way + host). 

\section{New Optical and Near-Infrared Observations}\label{sec:methods}

\input{obs-info}

Our new photometry includes pre-discovery non-detections and early $g$-band photometry from the All-Sky Automated Survey for SuperNovae \citep[ASAS-SN; ][]{shappee2014, kochanek2017} and post-maximum observations from the Transiting Exoplanet Survey Satellite \citep[TESS;][]{ricker2015}. New spectroscopic observations were acquired with the Gemini Multi-Object Spectrograph (GMOS; \citealp{GMOSref}) on the Gemini-North telescope and the Near-Infrared Echelle Spectrograph (NIRES, \citealp{NIRES1ref}) on the Keck~II telescope, a variation of the TripleSpec near-infrared (NIR) spectrograph \citep{NIRESref2, NIRESref3}. Basic information for the spectroscopic observations are provided in Table \ref{tab:obsinfo} and the new photometry of \name is provided in Table \ref{tab:newphot}.

\subsection{ASAS-SN Photometry}
\input{newphot}

New ASAS-SN $g$-band observations were reduced using a fully-automated pipeline \citep{shappee2014, kochanek2017} based on the ISIS image subtraction package \citep{alard1998, alard2000}. Each photometric epoch combines three dithered 90-second image exposures subtracted from a reference image. In addition to the standard pipeline we rebuilt the reference image excluding any images with $\rm{JD}\geq 2\,458\,812$ to prevent any flux contamination from the SN.

We then used the IRAF package \texttt{apphot} to perform aperture photometry with a 2-pixel, or approximately $16.\!\!''0$, radius aperture on each subtracted image, generating a differential light curve. The photometry was calibrated using the AAVSO Photometric All-Sky Survey \citep{henden2015}. All subtractions were inspected manually and images with clouds or other systematic issues are excluded from the final light curve.  

\subsection{TESS Photometry}

\name was observed by \textit{TESS} \citep{ricker2015} during the mission's Sector 21 and 22 operations, from Jan. 21.94 to March 17.96, 2020 UTC.
These observations narrowly miss the peak of the light curve, but provide excellent high-cadence monitoring of its subsequent decline.
\textit{TESS} observes in a single $\sim6\,000-10\,000~\rm\AA$ broad-band filter with an effective wavelength of $\sim8000~\rm\AA$ that is comparable to that of the Johnson-Cousins $I$-band.

We reduced the \textit{TESS} data using the image subtraction procedure of \cite{vallely2019b}, which implements a version of the ASAS-SN pipeline optimized for use with the \textit{TESS} Full-Frame Images (FFIs).
As in \cite{vallely2019b} and \cite{holoien2019}, due to the large pixel scale of the \textit{TESS} CCDs we chose to construct independent reference images for each sector rather than try to rotate a single reference image for use across multiple sectors.
For each sector, reference images were built using the first 100 FFIs of good quality, excluding those with sky background levels or PSF widths above average for the sector as well as those associated with mission-provided data quality flags.
The measured fluxes were converted into physical \textit{TESS}-band fluxes using an instrumental zero point of 20.44 electrons per second in the FFIs, based on the values provided in the TESS Instrument Handbook \citep{TESSHandbook}.
The flux offset between the two sector light curves was determined by using a linear extrapolation of the last 1.5 days of Sector 21 and the first 1.5 days of Sector 22.
Because the supernova had already attained maximum light prior to the first \textit{TESS} observations, the reference images necessarily include flux from the transient.
Precise calibration of the absolute flux is unimportant for our analysis, and for display purposes we simply normalize the \textit{TESS} photometry to the Zwicky Transient Facility \citep[ZTF; ][]{bellm2019} $i$-band photometry presented in \citet{miller2020}.

\subsection{GMOS Observations}

The data reduction process for the optical spectra, observed with GMOS on the Gemini-North telescope, generally follows the Gemini Data Reduction Cookbook\footnote{\url{http://ast.noao.edu/sites/default/files/GMOS_Cookbook/}}. Each night has observations at two separate grating angles to remove gaps from the final spectra. Raw frames are bias- and overscan-subtracted, flat-fielded, and mosiaced to reconstruct the monolithic detector. \textsc{lacosmic} \citep{dokkum2001} is used to detect and reject cosmic rays. Wavelength solutions are derived from arc-lamp exposures and the spectral response curves are generated from spectrophotometric standard stars for each grating tilt. After extracting each spectrum, the individual spectra from each night are combined into a single spectrum and the standard deviation of the individual spectra is used to estimate the uncertainty in the combined spectrum. 

We acquired $r$-band imaging on both nights to reliably flux-calibrate the GMOS spectra. These images are reduced using the same basic process as the spectroscopic observations including bias- and overscan-subtraction, flat-fielding, and cosmic-ray rejection. The World Coordinate System (WCS) solution provided by the Gemini \textsc{iraf} package is optimized with \textsc{astrometry.net} \citep{lang2010}. After the astrometric calibration procedure, we use $r$-band photometry from the Pan-STARRS survey \citep{chambers2016, flewelling2016} to calibrate each image. The first epoch of our GMOS observations overlaps with the end of the ZTF light curve \citep{miller2020}, confirming our spectrophotometry with a measured offset of $\Delta m = -0.005\pm0.007$~mag.

\subsection{NIRES Observations}\label{subec:methods.NIRES}

In addition to the optical GMOS observations, we observed \name with NIRES on the Keck~II telescope, which covers $0.95-2.45~\um$ with a resolution of $\lambda / \Delta \lambda \approx 2700$. The NIRES observations are normalized by the flat-field and A-B observation pairs are used to remove the sky background. \name is too faint to trace across all the echelle orders so the standard star observation is used as a trace template. Arc lamp exposures are used to derive the initial wavelength solution which is then improved with sky emission lines. The standard star observation is used to correct the spectrum for instrumental response and broadband atmospheric absorption. However, NIR photometry of \name could not be obtained before it set for the season, so the NIRES spectrum is presented on a relative flux scale.

\section{Photometric Comparisons}\label{sec:phot}

\subsection{Verifying the Distance to NGC~4441}

\cite{miller2020} adopted a distance of $42.5 \pm 2.1$~Mpc to NGC~4441 derived from the 2M++ peculiar velocity model \citep{carrick2015} which is inconsistent with the surface brightness fluctuation (SBF) distance of $\approx 19$~Mpc from \citet{tonry2001}. NGC~4441 is likely the merger of a spiral and an elliptical galaxy \citep{manthey2008} but SBF distance measurements are best for single-age stellar populations and can be skewed by spatial variations caused by dust or recent star-formation \citep{bothun1998} which may explain the difference between the kinematic and SBF distances. The SBF distance also implies a high peculiar velocity of $\sim 1\,300~\kms$ for NGC~4441 so we attempt to verify the validity of the kinematic distance by comparing the 2M++ results to the peculiar velocity field derived by \cite{graziani2019}. Their approach involves taking measured velocities and distances from Cosmicflows-3 \citep{tully2016} and applying a hierarchical Bayesian model to derive the peculiar velocity field out to $z\sim0.05$ \footnote{\url{http://edd.ifa.hawaii.edu/CF3calculator/}} \citep{kourkchi2020}. For NGC~4441, the model provides an expected distance of $D= 43.5$~Mpc, consistent within $\sim2\%$ of the 2M++ result of \cite{carrick2015}.

In addition to constraints from peculiar velocity reconstructions, we also check if NGC~4441 belongs to a nearby galaxy group. Based on the original SBF distance from \cite{tonry2001}, \cite{kourkchi2017} label NGC~4441 as a field galaxy with no other group members. Nearby in position and velocity, only one galaxy (NGC~4545) has a redshift-independent distance available in Cosmicflows-3, with a measured Tully-Fisher \citep{tully1977} distance of $D = 34.5 \pm 5.2$~Mpc and a preliminary Cosmicflows-4 measurement (Tully et al., in preparation) of $D = 36.6 \pm 7.3$~Mpc. New galaxy distances from Cosmicflows-4 also show several additional galaxies in the vicinity of NGC~4441 with Tully-Fisher distances ranging between 35 and 42 Mpc. The most likely scenario is that NGC~4441 is a member of a group with the nearby NGC~4521 as the brightest member. This reinforces the likelihood that the original SBF distance to NGC~4441 by \cite{tonry2001} is erroneous. Thus, we adopt the same distance to NGC~4441 as \citet{miller2020} for consistency, namely $D = 42.5 \pm 2.1$~Mpc and $\mu = 33.14\pm0.11$~mag.

\subsection{Confirming the Time of First Light}

\begin{figure}
    \centering
    \includegraphics[width=\linewidth]{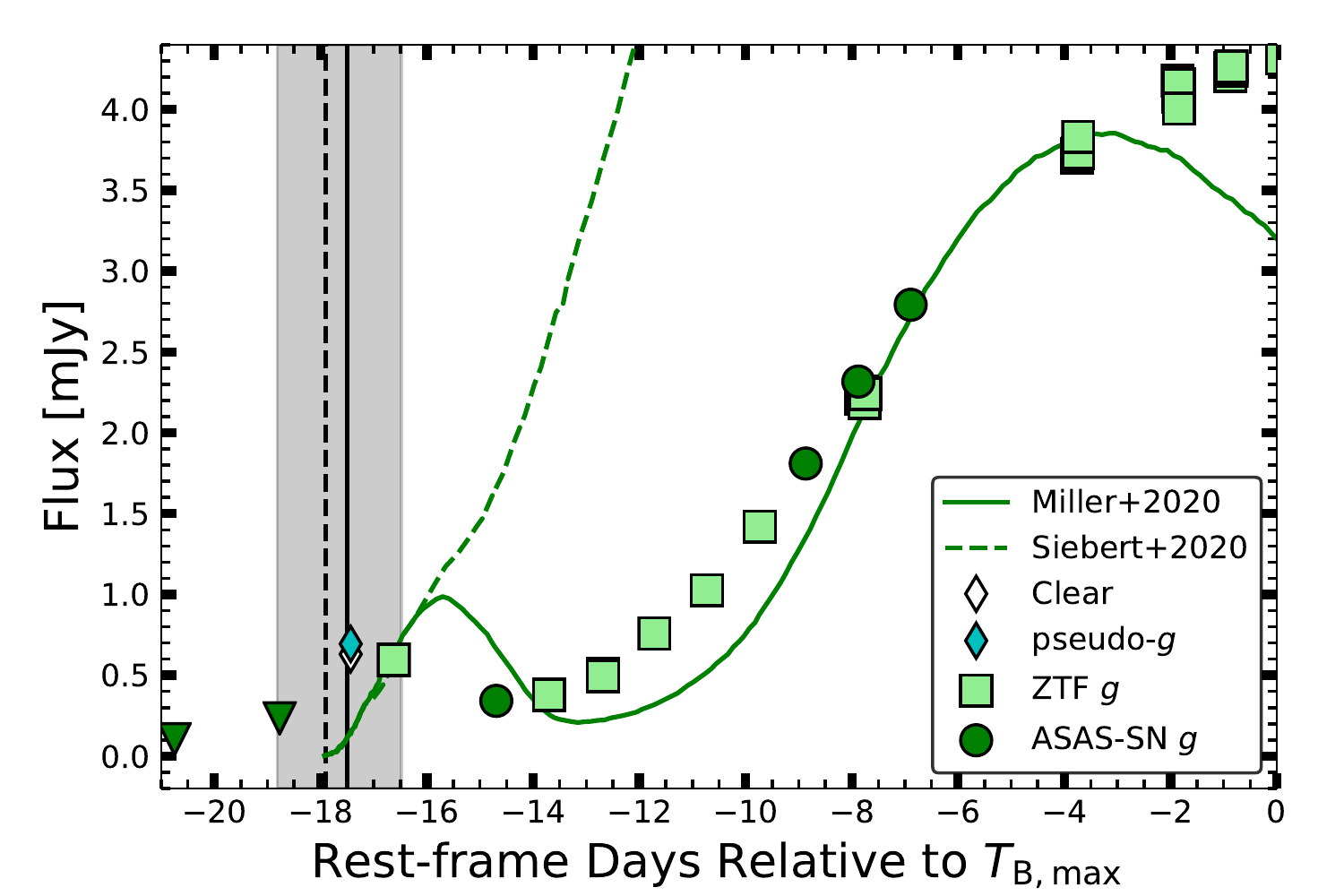}
    \caption{Early photometry of \name including our new ASAS-SN $g$-band detections and $3\sigma$ upper limits, ZTF $g$-band detections \citep{miller2020}, and the Clear-filter discovery measurement from \citet{itagaki2019}. Uncertainties for ASAS-SN and ZTF photometry are shown but they are usually smaller than the points. The \citet{itagaki2019} discovery magnitude did not include an uncertainty estimate. We compute a ``psuedo'' $g$-band flux measurement from the Clear-filter discovery magnitude with some assumptions (see text). The solid vertical line and shaded region is the photometric estimate of first light and the dashed black line is the first light time inferred from modeling the early spectra \citep{miller2020}. Solid and dashed green lines are double-detonation model light curves from \citet{miller2020} and \citet{siebert2020}, respectively. }
    \label{fig:earlyLC}
\end{figure}

Fig. \ref{fig:earlyLC} shows the early $g$-band light curve including our new photometry, the ZTF $g$-band observations from \citet{miller2020}, and the discovery magnitude from \citet{itagaki2019} compared to the double-detonation light curve models from \citet{miller2020} and \citet{siebert2020}. The photometry from \citet{itagaki2019} was obtained with a Clear filter which we convert to an approximate $g$-band magnitude for easier comparison. This conversion assumes the Clear filter can be approximated as a combination of $g$ and $r$ filters, that the color evolution is negligible between the discovery observation and the first ZTF observations (i.e., a constant color of $g-r\approx -0.2~$mag), and a conversion between AB and Vega systems of $\sim 0.2$~mag similar to comparable optical filters. As the early spectra rapidly evolve from a strong blue continuum \citep{miller2020}, this likely slightly underestimates the true $g$-band flux. The pseudo-$g$ magnitude is included only for instructive purposes and we caution that the associated uncertainties are likely of order $20\%$.

The time-of-discovery from \citet{itagaki2019} coupled with the ASAS-SN non-detection $\approx 1.3$~days prior to discovery constrain the likely first-light time $t_{fl}$ to MJD $58\,844.4-58\,845.7$ (18.8 to 17.4~days before $T_{\rm{B,max}}$). This is mostly consistent with the photometric estimate from \citet{miller2020} and confirms their decision to exclude the two early ZTF observations when fitting $t_{fl}$, although their $t_{fl}$ is likely too late (i.e., too close to the \citealp{itagaki2019} discovery) to be physically reasonable. The last ASAS-SN non-detection requires a minimum rise of $\gtrsim 8.0~\mu$Jy/hour whereas the photometric $t_{fl}$ from \citet{miller2020} implies a rise of $\sim 400~\mu$Jy/hour. Even if the \citet{itagaki2019} measurement is erroneous by over a magnitude, using $g = 18$~mag still implies a rise of $\sim 150~\mu$Jy/hour for the photometric $t_{fl}$ from \citet{miller2020}. Thus, the spectroscopic rise time of $\sim 18$~days from \citet{miller2020} is likely closer to the true rise time and is consistent with our new non-detections and early photometry. 


\begin{figure*}
    \centering
    \includegraphics[width=0.45\linewidth]{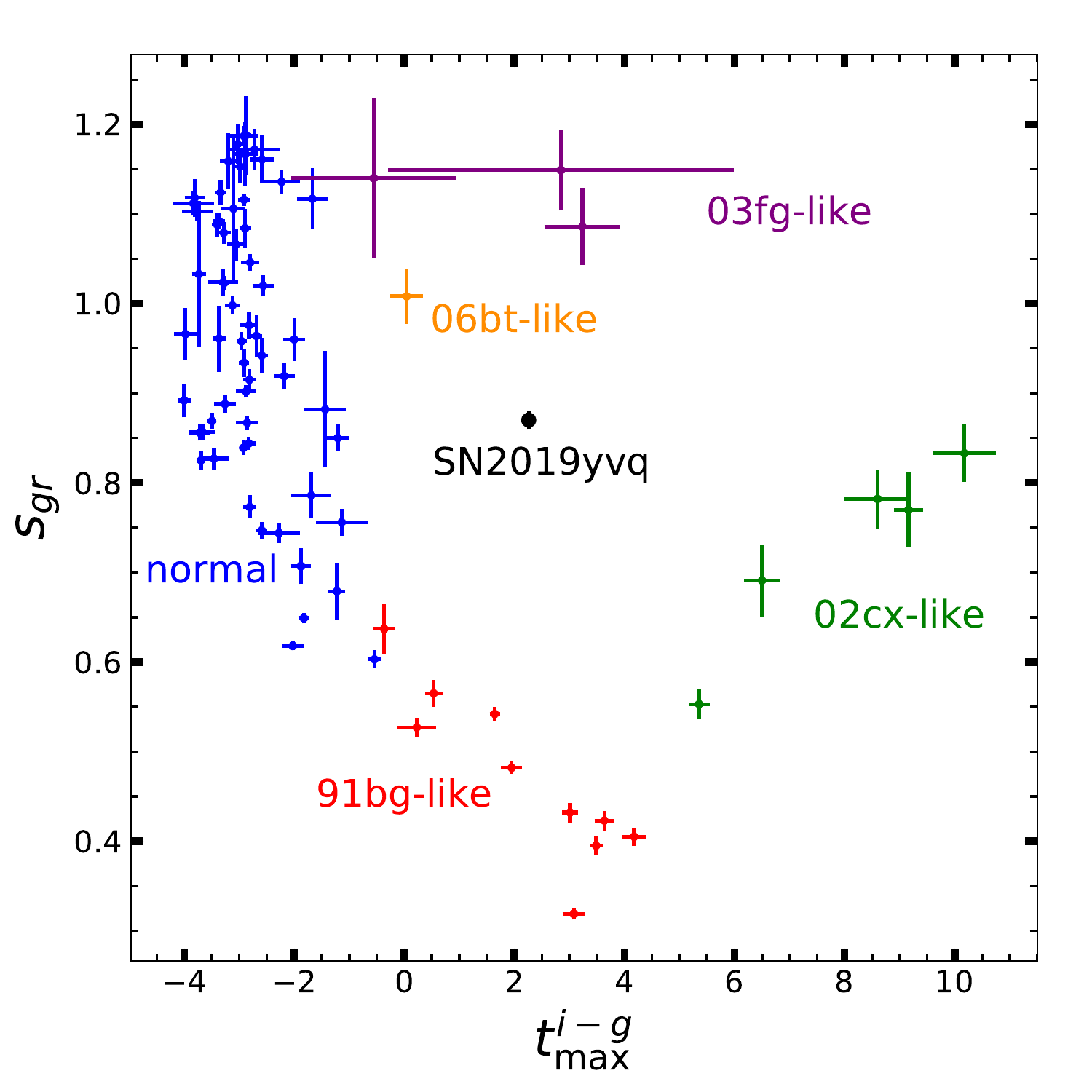}
    \includegraphics[width=0.45\linewidth]{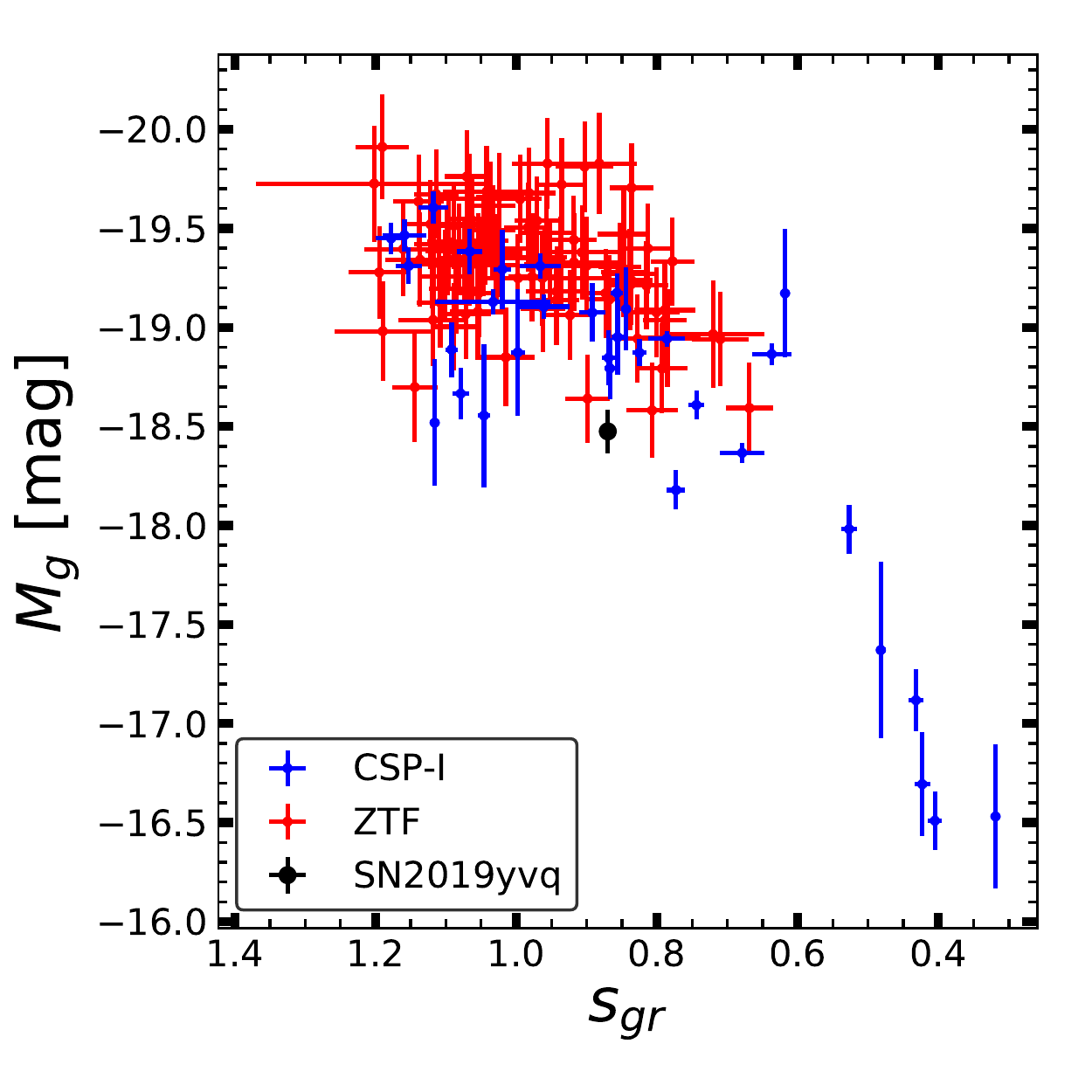}
    \caption{
    \textit{Left}: \name compared to the photometric classification scheme from \citet{ashall2020}.
    \textit{Right}: Stretch versus peak $g$-band absolute magnitude for \name relative to \sne from CSP-I \citep{krisciunas2017} and ZTF \citep{yao2019}.
    }
    \label{fig:sgr}
\end{figure*}

\subsection{Near-Peak Comparisons}

We compare \name to the photometric properties of other \sne near maximum light in Fig. \ref{fig:sgr}. \name is not well-fit by standard \sn light curve models \citep{miller2020} so we fit polynomials to measure the light curve parameters and employ bootstrap-resampling to estimate the associated uncertainties. Fitting quadratic polynomials to the ZTF $g$- and $i$-band light curves near peak we find an offset of $t_{\rm{max}}^{i-g} = 2.26\pm0.08~\rm{days}$ and $g_{\rm{max}} = 14.82\pm0.01$~mag for \name.

Another metric for \sne light curves is the stretch $s_{XY}$ \citep{burns2014} measured from two filters $X$ and $Y$ which utilizes the color evolution instead of the decline rate to standardize light curves (e.g., $s_{BV}$ and $s_{gr}$; \citealp{ashall2020}). We use spline fits to resample the ZTF $r$-band light curve \citep{miller2020} and associated uncertainties to the epochs of the $g$-band light curve. Then, we fit a quadratic polynomial to the $g-r$ light curve and derive $s_{gr} = 0.87 \pm 0.01$. These light curve parameters place \name in an unoccupied region of parameter space in the left panel of Fig. \ref{fig:sgr}, reinforcing both the uniqueness and rarity of \name-like events.  

The right panel of Fig. \ref{fig:sgr} compares the $g$-band absolute magnitude and stretch of \name to \sne from the Carnegie Supernova Project I \citep[CSP-I;][]{krisciunas2017} and the ZTF 2018 sample \citep{yao2019}. $s_{gr}$ is measured directly from the $g$- and $r$-band light curves for the CSP \sne. The $g$- and $r$-band light curves for the ZTF sample \citep{yao2019} often do not extend to $\gtrsim30$~days after maximum so we fit these \sne with templates in SNooPy \citep{burns2011} which computes a template-derived $s_{BV}$. We then convert $s_{BV}$ to $s_{gr}$ using the relation from \citet{ashall2020}. Due to having observations in only two filters, the ZTF objects have a significant uncertainty due to the limited constraints on the host-galaxy reddening. CSP and ZTF observe in slightly different $g$-band filters with an offset of $g_{\rm{CSP}} - g_{\rm{ZTF}} \approx -0.03$~mag for normal \sne at peak-light up to $z < 0.1$ and we correct for this offset in our comparison. \name is fainter than \textit{all} \sne within $s_{gr} \pm0.1$ and fainter than all normal \sne in the ZTF 2018 sample (cf. Fig 6 from \citealp{miller2020}).

\subsection{Post-Maximum Comparisons}\label{subsec:phot.latephot}

\begin{figure}
    \centering
    \includegraphics[width=\linewidth]{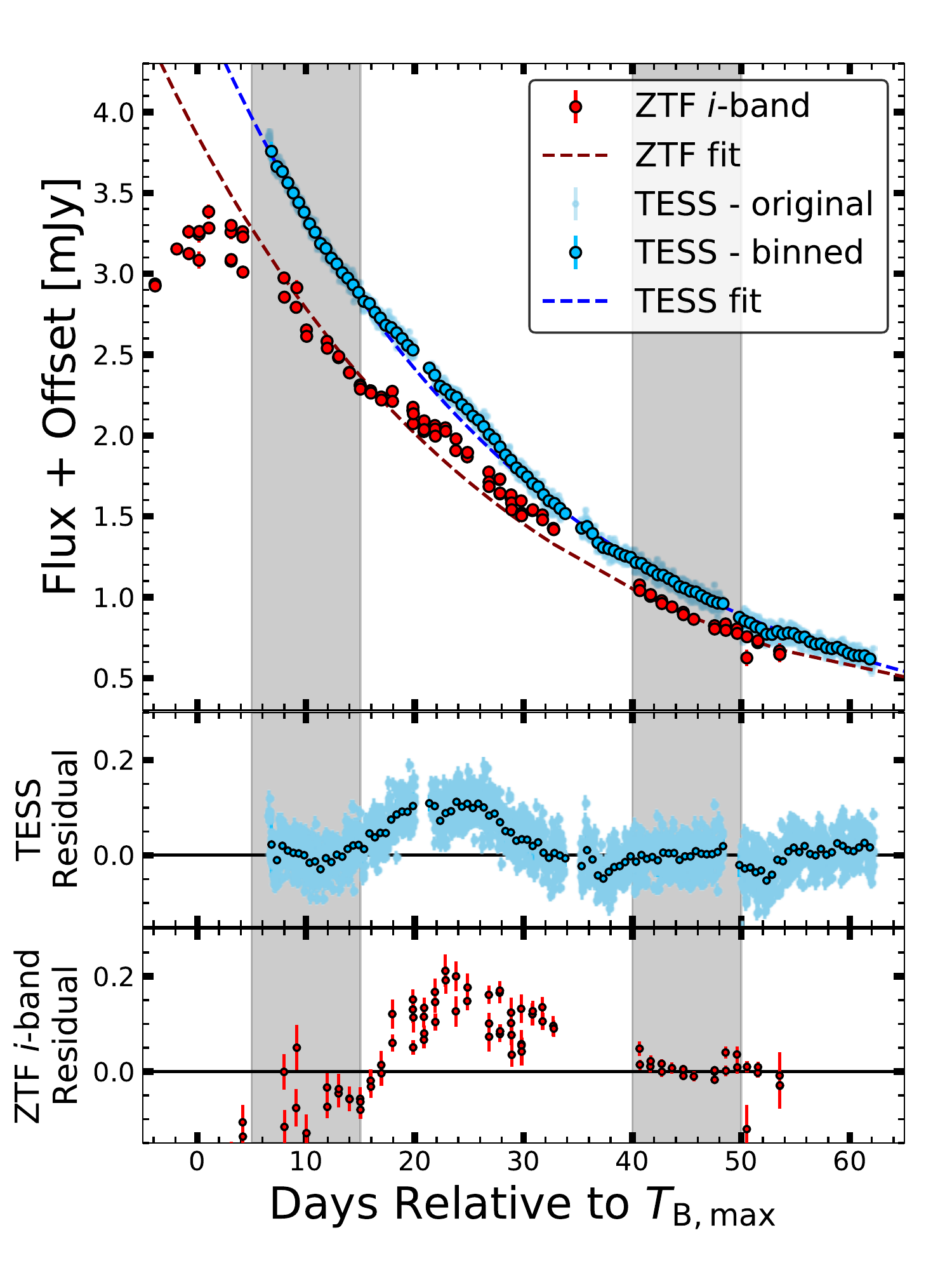}
    \caption{Post-maximum light curves of \name including the new TESS light curve (blue) and the ZTF $i$-band light curve (red) from \citet{miller2020}. The TESS data are offset by $+0.5$~mJy in the top panel to highlight the weak excess at $\sim 23$~days after maximum light. The TESS data are binned at 12~hour increments for visual clarity (dark blue). Exponential fits to the ZTF and binned TESS light curves are provided as dashed lines with the gray-shaded regions marking epochs used in fitting the exponential decay. The middle and bottom panels provide the residuals from the exponential decay fits for TESS and ZTF, respectively, revealing a very weak ``bump'' in both filters.}
    \label{fig:TESS}
\end{figure}

\name also exhibits unique properties after maximum light, in particular the lack of a prominent NIR secondary maximum \citep{miller2020}. Fig \ref{fig:TESS} shows the post-maximum TESS and ZTF $i$-band \citep{miller2020} light curves. The TESS light curve shows a slight inflection at a time that roughly coincides with the NIR secondary maximum of normal \sne \citep[e.g., ][]{kasen2006, ashall2020}. Only 91bg-like, 02cx-like and ``super-Ch"-mass (03fg-like) \sne lack prominent secondary maxima \citep[e.g., ][]{gonzalez2014, ashall2020} but \name is spectroscopically inconsistent with these sub-types of \sne \citep{miller2020}. The timing and presence of the NIR secondary maxima are thought to be caused by the recombination of Fe-group elements in the ejecta \citep{hoeflich2002, kasen2006, jack2015}. This can be seen in brighter \sne which have stronger and later NIR secondary maxima whereas dimmer \sne have no NIR secondary maximum \citep[e.g., ][]{taubenberger2017}. Similar \sne having intermediate luminosities but a weak/absent NIR secondary maximum include SN~2006bt \citep{foley2010}, SN~2002es \citep{ganeshalingam2012}, and SN~2006ot \citep{krisciunas2017}. However, these \sne have low photospheric velocities which is at odds with the high photospheric velocities observed in \name \citep{miller2020}. 

\section{Nebular Spectroscopy}\label{sec:spec}

\begin{figure*}
    \centering
    \includegraphics[width=\linewidth]{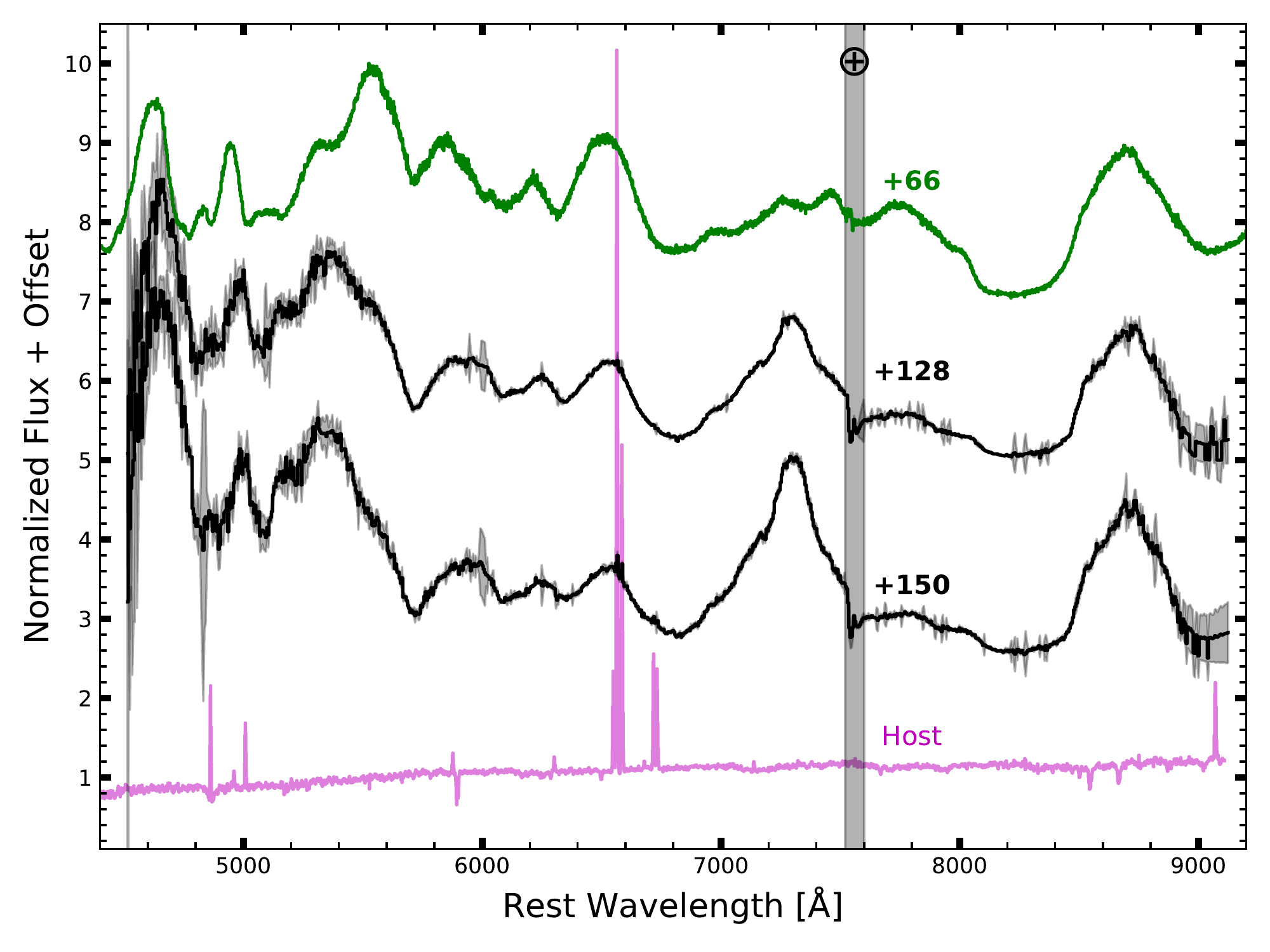}
    \caption{Late-phase and nebular spectra of \name (black), +66~day spectrum (green) from \citet{miller2020}, and the host spectrum (magenta) from SDSS \citep{york2000, abolfathi2018}. Uncertainties are given in gray, and regions of higher uncertainty are caused by detector chip gaps or sky emission lines. }
    \label{fig:optspex}
\end{figure*}

\begin{figure*}
    \centering
    \includegraphics[width=\linewidth]{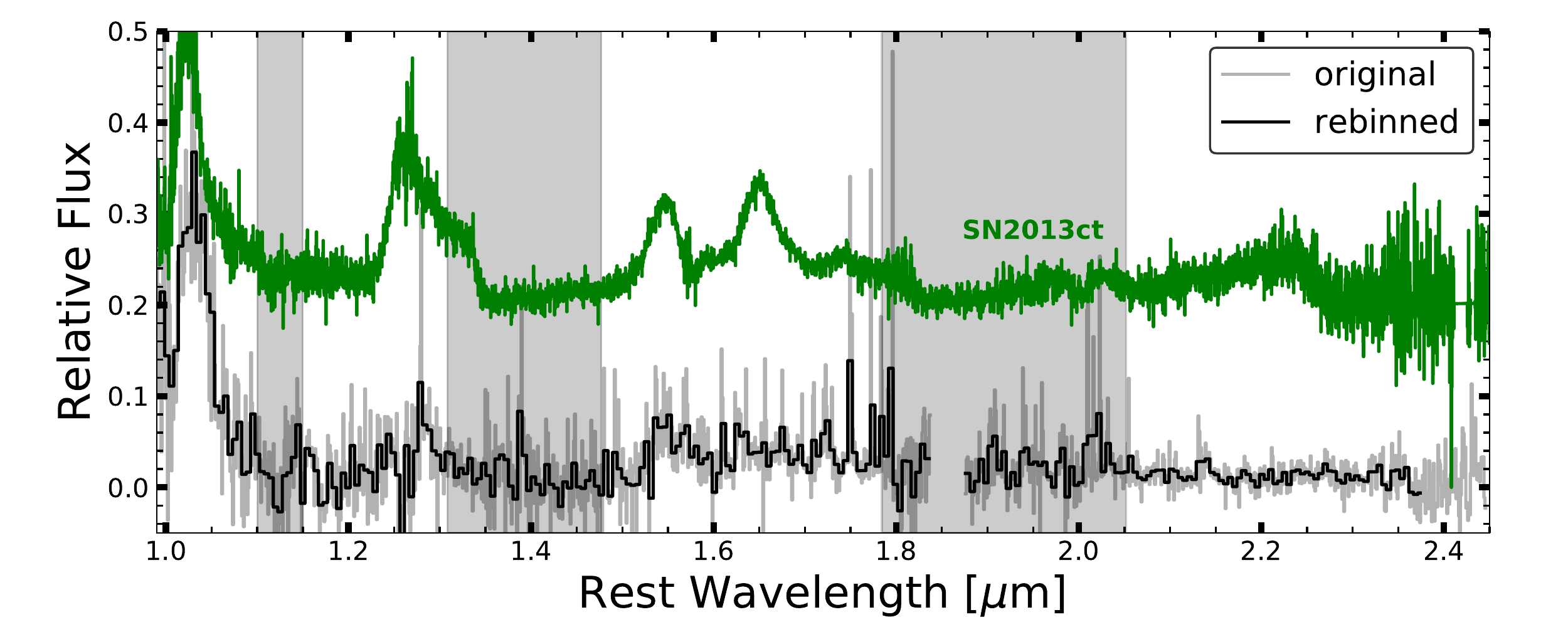}
    \caption{NIRES spectrum of \name at +173 days after maximum light and a +229~day spectrum of SN~2013ct (green) provided for comparison \citep{maguire2016}.}
    \label{fig:NIRESfull}
\end{figure*}

\begin{figure}
    \centering
    \includegraphics[width=\linewidth]{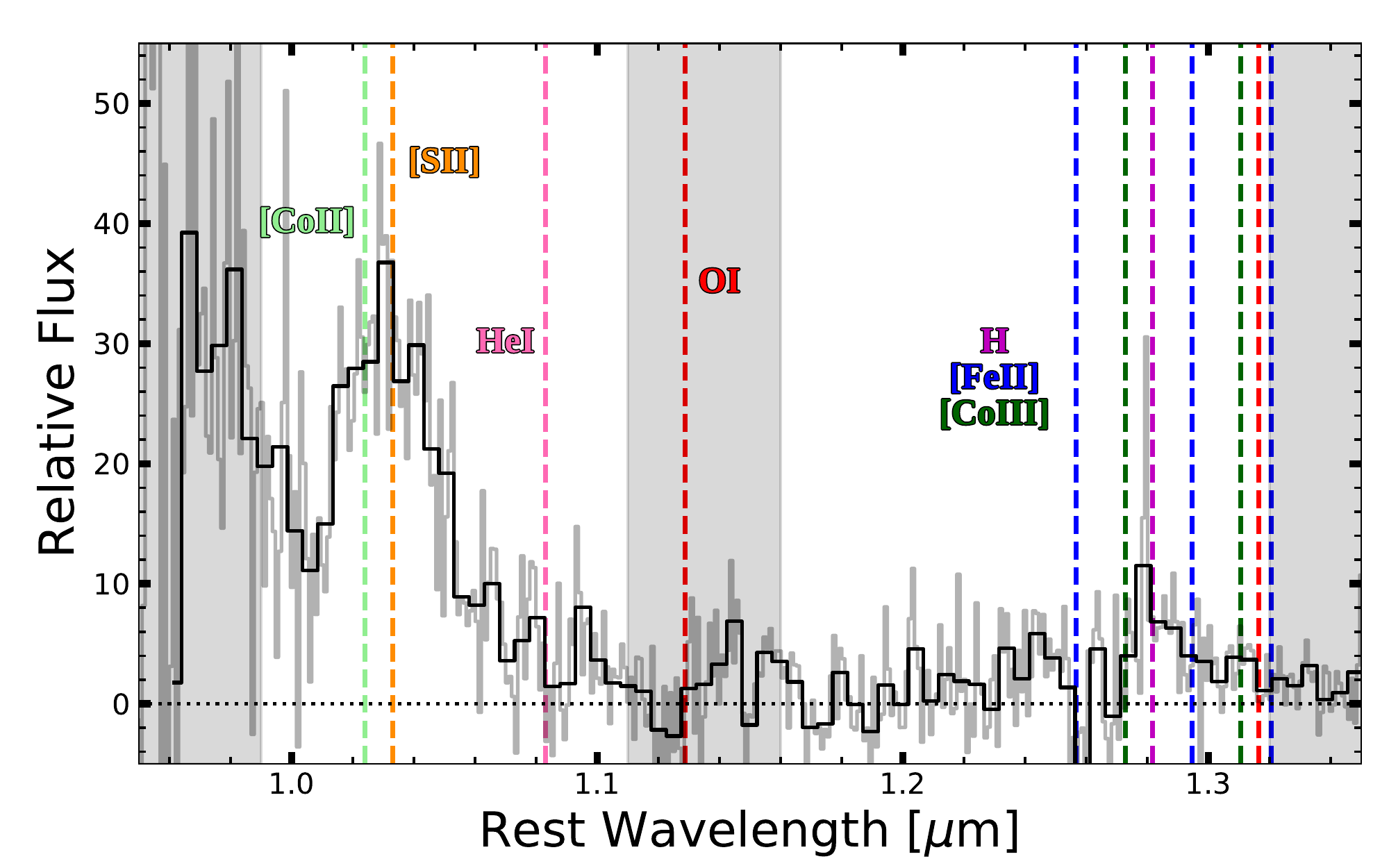}
    \caption{Subsection of the NIR spectrum with colored vertical lines marking the wavelengths of potential emission lines.}
    \label{fig:NIRESzoom}
\end{figure}

The late-phase optical spectra are shown in Fig. \ref{fig:optspex} and the nebular NIR spectrum is provided in Fig. \ref{fig:NIRESfull}. Fig. \ref{fig:NIRESzoom} shows expanded and labeled regions of the NIR spectrum uncontaminated by night sky lines. The optical spectra have many similarities to normal \sne such as strong emission lines of Ni, Co, and Fe. However, there are important differences which provide unique constraints on the progenitor system and explosion mechanism of \name. 

\subsection{Non-Detections of Stripped Companion and Merger Remnant Material}\label{subsec:spec.stripmaterial}

\begin{figure*}
    \centering
    \includegraphics[width=\linewidth]{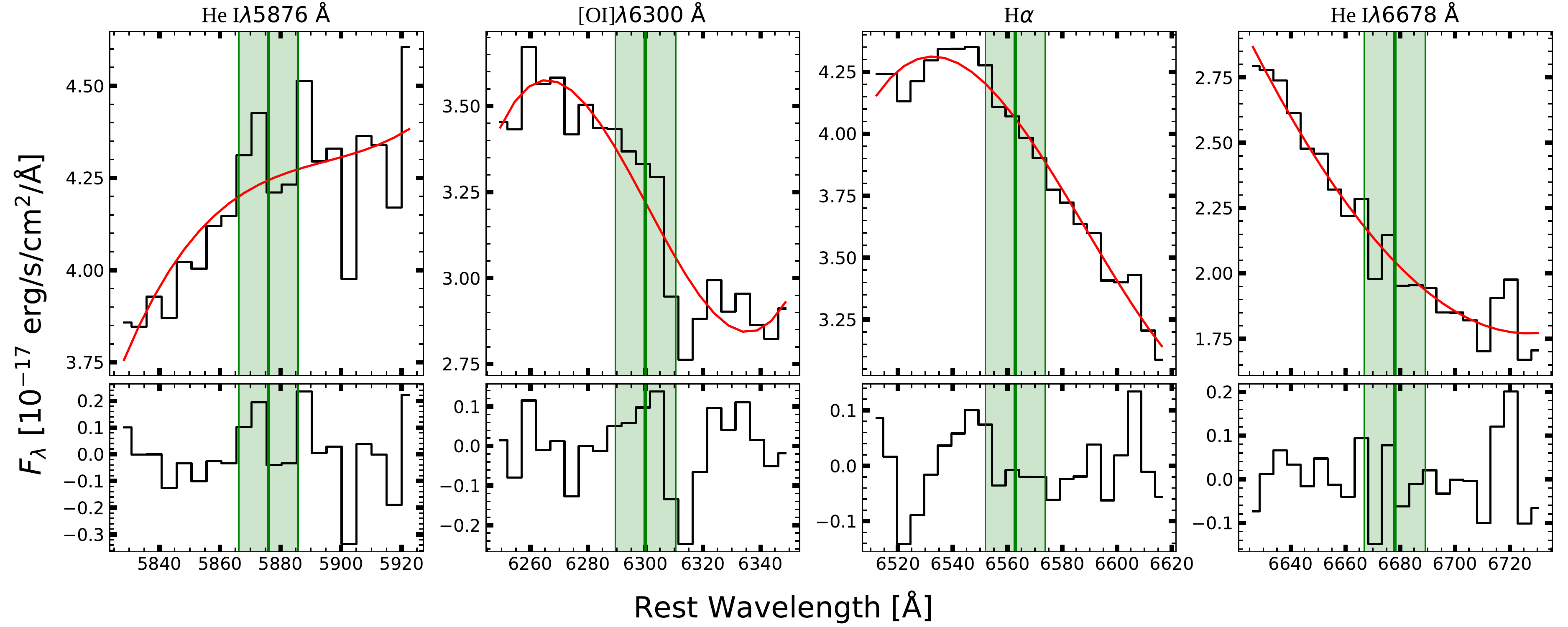}
    \caption{Non-detections for \ion{He}{1}, [\ion{O}{1}], and \Ha in the +150~d GMOS spectrum. The top panels show the observed spectrum + continuum model and the bottom panels show the continuum-subtracted spectrum. Green-shaded regions span the line FWHM of $\sim 1000~\kms$ expected for material stripped from a nearby donor star \citep[e.g., ][]{boehner2017}.}
    \label{fig:nondetect}
\end{figure*}

 We do not detect any H or He emission lines in our optical and NIR spectra, including \Ha, \Pa, \Pb, \ion{He}{1}$\lambda5876,6678\rm\AA$, and \ion{He}{1}$\lambda1.083,2.059\um$. Several previous studies place limits on non-detections on H/He in nebular spectra \citep[e.g., ][]{sand2019, tucker2020} relying on radiative-transfer models to predict the observed emission \citep[e.g., ][]{mattila2005, botyanszki2018}. However, \citet{dessart2020} model the effect of changing optical depth on the visibility of the H and He emission and find a non-negligible time-dependent impact from line-blanketing, especially on the higher-order Balmer lines. Thus, we place upper limits on the equivalent width $W$ following the procedure of \citet{leonard2007} (also see \citealp{leonard2001}). For both epochs of optical spectra, the equivalent width limit on \Ha emission of $W_{\Ha}(3\sigma) < 0.6~\rm\AA$ excludes any reasonable non-degenerate companion \citep{dessart2020}. This precludes the need for scaling the spectra to later epochs and assuming a non-variable spectral shape to match the nebular spectra models of \citet{botyanszki2018} as done by \citet{siebert2020}.
 
 We also do not detect permitted or forbidden \ion{O}{1} in the NIR (Fig. \ref{fig:NIRESzoom}) or optical spectra (Fig. \ref{fig:nondetect}). Oxygen emission is expected for violent mergers of two C/O WDs as the secondary (lower-mass) WD is only partially burnt \citep{pakmor2012}. SN~2010lp \citep{taubenberger2013} exhibited strong [\ion{O}{1}] emission in its nebular spectra and a violent merger is the preferred explanation for this feature \citep{taubenberger2013, kromer2013}. We note that O emission is only expected for merging C/O WDs and may not be present in the merger of a C/O WD and a He WD which we discuss further in \S\ref{subsec:discuss.CSM}.

\subsection{The \CaII Feature}

\begin{figure}
    \centering
    \includegraphics[width=1.02\linewidth]{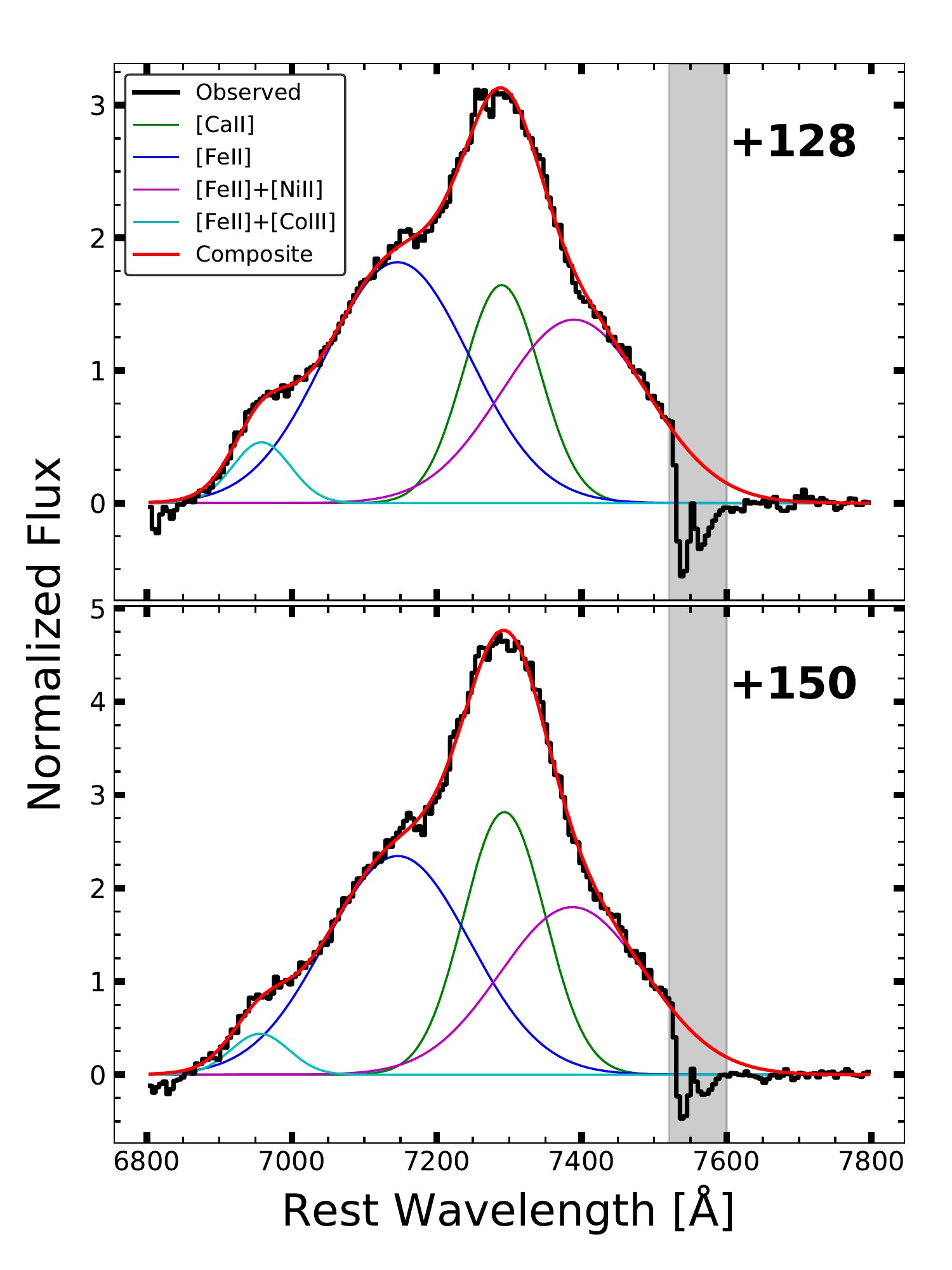}
    \caption{Fits to the \CaII region. The gray-shaded area marks a region of telluric contamination which is excluded from the fitting process.}
    \label{fig:CaIIfit}
\end{figure}

The sharp spectral feature at $\sim 7300~\rm\AA$ is almost certainly \CaII emission \citep{siebert2020} although \CoII, \FeII, and \NiII contribute to its wings. We lack the spectral resolution for a full decomposition of this region as done by \citet{siebert2020} and instead focus on the temporal evolution of the \CaII emission between our two spectra since this has not been examined previously. We assume the \CaII emission and its wings can be approximated by Gaussian profiles even though they are blends of multiple lines \citep[e.g., ][]{mazzali2015}. We require the velocity shifts and widths to be roughly consistent between the two epochs, allowing for a slight red-shifting and widening of the profiles in the later spectrum to account for the decreasing opacity of the ejecta \cite[e.g., ][]{black2016}.

The results are shown in Fig. \ref{fig:CaIIfit}. The absolute \CaII flux decreases between the two epochs but increases relative to the nearby \NiII and \FeII emission lines. Interestingly, the \CaII feature evolves differently than the surrounding \FeII + \NiII features and the strong emission line at $\sim 8700~\rm\AA$, but similarly to the \FeIII emission line at $\sim 4600~\rm\AA$ with a near-constant \CaII/\FeIII flux ratio of $\sim 0.4$.

\subsection{Origin of the \rm{8700~\AA{}} \textit{Emission Line}}

\begin{figure}
    \centering
    \includegraphics[width=\linewidth]{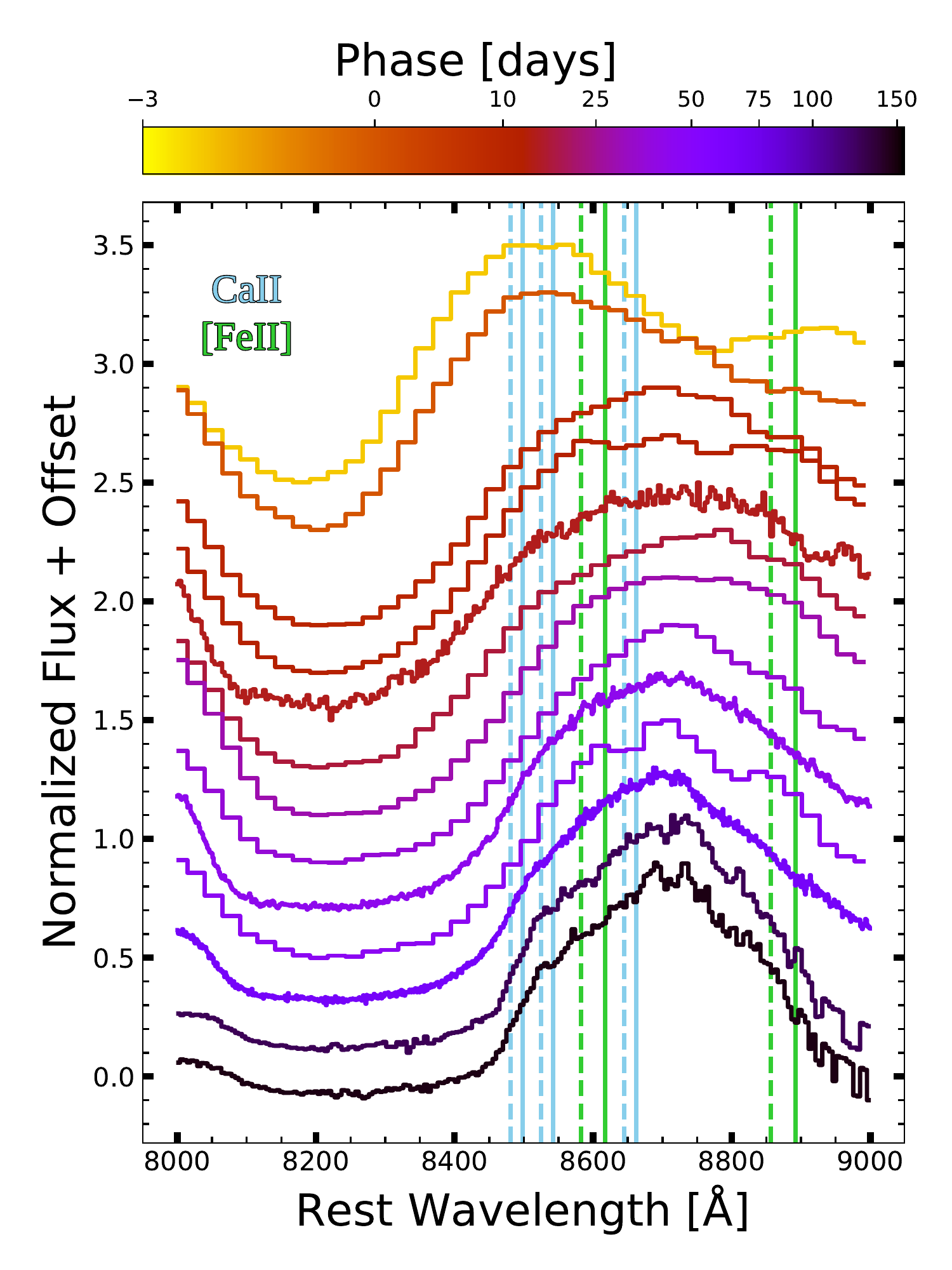}
    \caption{Evolution of the \CaIINIR NIR triplet including pre- and post-maximum spectra from \citet{miller2020} and the last two spectra being our new observations. Solid lines mark the rest wavelengths of the \CaIINIR NIR triplet (blue) and the \FeII $\lambda8617,8892~\rm\AA$ (green) emission lines. Dashed lines are the locations of \CaIINIR and \FeII shifted by $-600~\kms$ and $-1\,200~\kms$, respectively \citep{siebert2020}.}
    \label{fig:CaIINIRevol}
\end{figure}

The strong emission feature at $\sim8700~\rm\AA$ is uncommon in nebular spectra of normal \sne but is usually seen in near-maximum spectra and attributed to absorption and emission of the permitted \CaIINIR triplet \citep[e.g., ][]{branch2006, branch2008}. However, this feature usually disappears by $\lesssim 100$~days after maximum light (e.g., $\lesssim75$~days for SN~2011fe, \citealp{pereira2013}; $\lesssim 90$~days for SN~1991bg, \citealp{turatto1996}; $\lesssim 80$~days for SN~1991T, \citealp{silverman2012}; also see \citealp{branch2008}) yet the feature is essentially unchanged between near-maximum spectra and our two spectral epochs at +128 and +150~days (Fig. \ref{fig:CaIINIRevol}). \citet{siebert2020} attribute this feature to \CaIINIR and the smooth evolution of this feature from maximum light through our observation reinforces this conclusion. 

\begin{figure}
    \centering
    \includegraphics[width=\linewidth]{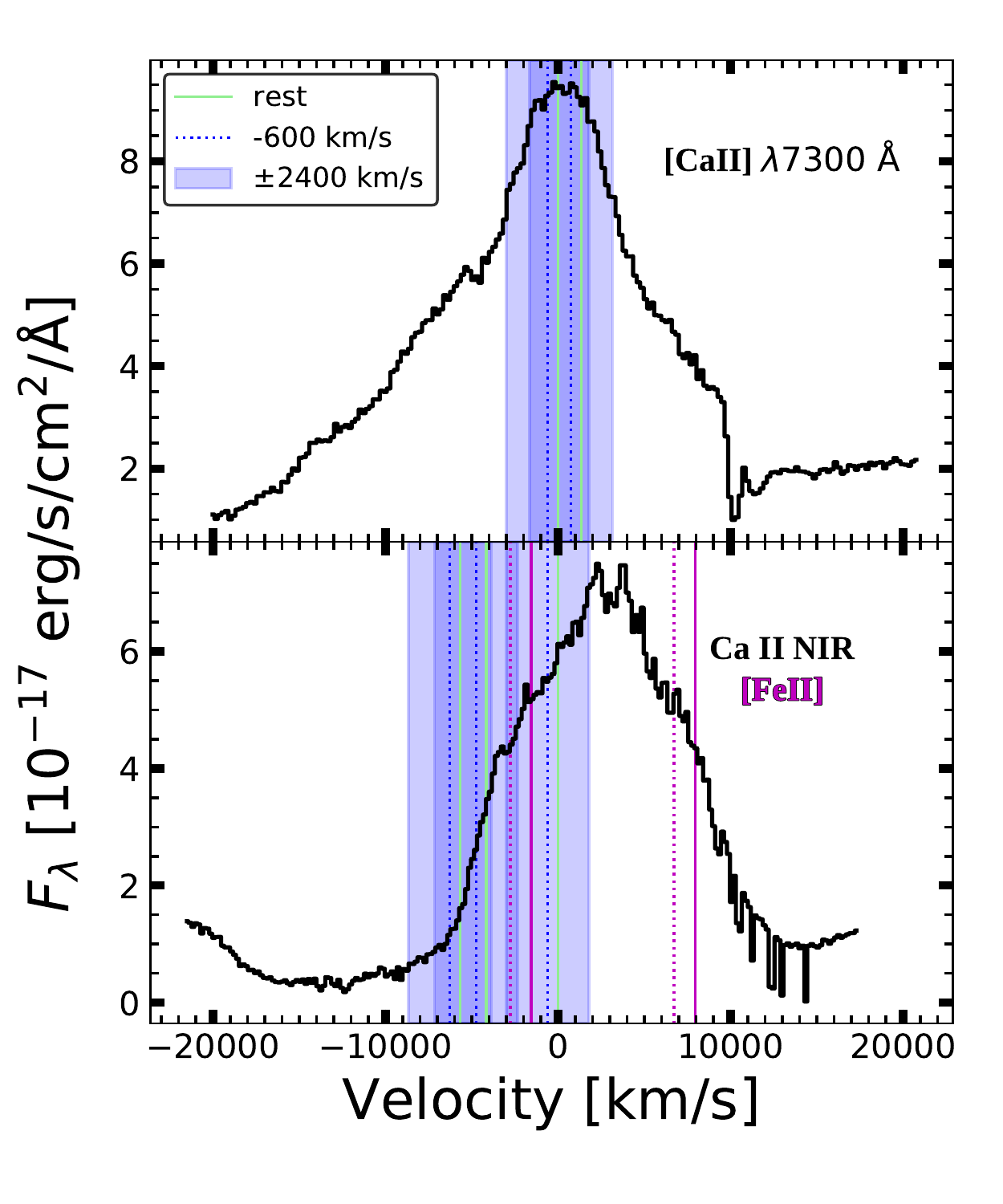}
    \caption{A velcoity-space comparison of the \CaII (top) and \CaIINIR NIR (bottom) emission for the $+150$~day spectrum. The top panel is centered on the stronger \CaII emission component ($\lambda 7291~\rm\AA$) and the bottom panel is centered on the reddest component of the \CaIINIR NIR triplet ($\lambda8662~\rm\AA$). The rest wavelengths of each Ca line are shown in green and the derived velocity shift and width of the \CaII feature derived by \citet{siebert2020} are provided in blue. Only the central narrow component of the top panel originates from \CaII (see Fig. \ref{fig:CaIIfit}) whereas the majority of the emission profile in the lower panel is likely dominated by the \CaIINIR NIR triplet (i.e., Fig. \ref{fig:CaIINIRevol}). The purple lines in the bottom panel mark \FeII$\lambda8617,8892~\rm\AA$ at its rest wavelength (solid) and shifted by $-1200~\kms$ (dotted, \citealp{siebert2020}).}
    \label{fig:CaIIcompare}
\end{figure}

The \CaIIlong and the \CaIINIR NIR emission profiles are juxtaposed in Fig. \ref{fig:CaIIcompare}. The irregularities in the \CaIINIR NIR emission profile at the velocity shifts for \CaIIlong and \FeII derived by \citet{siebert2019} suggest that the Ca-emitting material producing the \CaIIlong feature does contribute to the \CaIINIR NIR feature but cannot explain the entire emission profile. There is a clear absorption trough centered at $\sim 8300~\rm \AA$ (Figs. \ref{fig:optspex} and \ref{fig:CaIINIRevol}), so this profile is likely a complex blend of absorption and emission. Additionally, there are other lines present at this location typical of nebular \sne such as \FeII$\lambda8617,8892~\rm\AA$. However, the Fe transitions in the $\lambda4500-5000\rm\AA$ are roughly consistent with normal \sne \citep{siebert2020} so this explanation for the entire $\lambda8700\rm\AA$ feature is unlikely. Thus, it is likely that this emission is dominated by the \CaIINIR NIR triplet but that it does not originate in the same region of the ejecta as the \CaIIlong emission.

The absorption component of the \CaIINIR NIR triplet is a common feature of \sne near maximum light which transitions into a complex blend of absorption and emission features in the weeks after maximum light \citep[e.g., ][]{maguire2014, silverman2015, siebert2019}. The absorption component commonly has different polarization angles and strengths than the surrounding continua \citep[e.g., ][]{wang2003, leonard2005, wang2006} suggesting an asymmetric distribution of absorbing material outside the photosphere \citep[e.g., ][]{cikota2019}. Additionally, there is a subset of \sne with high-velocity absorption features (e.g., \citealp{mazzali2005}) where the high-velocity component of the \CaIINIR NIR triplet is kinematically distinct from the photospheric component and thus likely originates in a different portion of the ejecta \citep[e.g., ][]{wang2003, leonard2005, wang2009}. Considering the high \SiII velocities of \name \citep{miller2020}, it is possible that this persistent \CaIINIR emission is related to high-velocity \sne and would naturally explain why the kinematic parameters of the \CaIIlong and \CaIINIR NIR triplet are dissimilar. However, other \sne with high-velocity absorption features lack a prominent \CaIINIR NIR emission component at $>100$~days after maximum (e.g., SN~2002dj, \citealp{pignata2008}; SN~2009ig, \citealp{foley09ig}). Thus, it is unclear if these similarities are coincidental or if the differences are due to intrinsic variations of a common source such as line-of-sight variations, different CSM configurations, or some other aspect of the progenitor system and/or explosion mechanism.

\subsection{NIR Spectrum}

The nebular-phase NIRES spectrum is shown in Fig. \ref{fig:NIRESfull} and possible emission lines are marked in Fig. \ref{fig:NIRESzoom}. Consistent with our optical spectra, we do not detect the expected emission lines from any potential stripped companion material such as \Pa or \ion{He}{1}$\lambda1.083~\mu\rm m$. Stripped material models predict the NIR H and He emission lines to be the strongest features in the spectrum \citep{maeda2014, botyanszki2018, dessart2020} and our non-detections severely restrict any potential stripped mass or nearby CSM.

The only major detected feature occurs at $\sim 1.03~\mu\rm m$ which we attribute to a blend of \CoIII and \SII. \citet{siebert2020} derived velocity shifts for the $^{56}$Ni decay products on order of $\sim -4\,000~\kms$, so this feature is likely dominated by \SII since a \CoII would need to be red-shifted. We lack the signal-to-noise to decompose the individual contributions. This feature is stronger than any other feature in the NIR spectrum by a factor of $\approx 3$. 

The $1.15-1.4~\um$ region exhibits little flux with the exception of a feature at $\approx 1.27~\um$. This coincides with the location of \Pa, but it is unlikely to stem from stripped companion material for several reasons. The host spectrum from SDSS (Fig. \ref{fig:optspex}, \citealp{york2000, abolfathi2018}) has strong \Ha emission indicating star-formation which is generally accompanied by Paschen emission in NIR spectra \citep[e.g., ][]{hill1996}. Additionally, the emission profile is narrower than expected for stripped companion material \citep[e.g., ][]{boehner2017, botyanszki2018} but has a velocity width consistent with the optical host-galaxy emission lines \citep[e.g., ][]{fernandes2005}. The broader component is likely a blend of \CoIII and \FeII, as our epoch (+173~days after maximum light) is when the NIR spectrum is transitioning from Co-dominated to Fe-dominated \citep[e.g., ][]{sand2016,diamond2018}.

The $1.5-1.8~\um$ region has non-zero flux which can be attributed to a complex blend of \FeII, \FeIII, \CoIII, and \NiIII \citep{diamond2015, maguire2016, diamond2018}. Due to the low signal-to-noise of this region we cannot model the various contributing emission components. 


\section{Discussion}\label{sec:discuss}

There are several possible explanations for the early UV/optical flux excess seen in \name, such as interaction with a nearby companion star \citep{kasen2010}, mixing of $^{56}$Ni to the outer layers of the ejecta \citep{piro2012}, a double-detonation driven by an accreted surface He layer \citep{khokhlov1993}, or a violent merger of two WDs \citep{pakmor2010,pakmor2012}. Each scenario has predicted consequences for both near-peak and nebular-phase observables and we compare our new data to these expectations. 

\name is unique in its near-peak photometric properties but exhibits an absolute luminosity mostly consistent with \sne (Fig. \ref{fig:sgr}). The nebular spectra of \name exhibit many of the features seen in normal \sne such as broad emission lines of Co and Fe (Figs. \ref{fig:optspex} and \ref{fig:NIRESzoom}). The optical and NIR spectra have no indication of H/He emission lines (Figs. \ref{fig:NIRESzoom} and \ref{fig:nondetect}) expected for material stripped/ablated from a nearby companion star \citep[e.g., ][]{botyanszki2018, dessart2020} nor [\ion{O}{1}] emission expected from a violent merger of two C/O WDs \citep[e.g., ][]{taubenberger2013, kromer2013}. However, there are two features of interest not typically seen in late-phase \sne: \CaIIlong emission and the persistence of the \CaIINIR NIR triplet \citep{siebert2020}.

The \CaII feature increases in strength between the nebular-phase optical spectra at +128 and +150~days after $T_{\rm{B,max}}$. This increase is not seen in the nearby Co or Ni emission lines but is reflected in the \FeIII$\lambda4660\rm\AA$ emission line with a near constant \CaII/\FeIII ratio of $\approx 0.4$. This could be due to the decreasing optical depth of the ejecta, which plays an important role in the strength and visibility of emission lines at these epochs \citep[e.g., ][]{dessart2020} and is consistent with Ca-emitting material residing near the center of the explosion.

The \CaIINIR NIR triplet requires more modeling to completely understand. The \CaIINIR emission has irregularities in the emission profile coinciding with the blue-shift of the \CaIIlong feature measured by \citet{siebert2020}, indicating that some of this emission is powered by the same source. However, most of the \CaIINIR triplet is inconsistent with the \CaII velocity profile and the discrepancy cannot be explained by the presence of \FeII emission lines (Fig. \ref{fig:CaIIcompare}). Additionally, there seems to be no temporal evolution of the \CaIINIR NIR triplet between our two optical epochs in either the relative emission line strengths or the overall profile shape (Fig. \ref{fig:CaIINIRevol}). This suggests that the Ca-rich material responsible for the higher-velocity \CaIINIR NIR triplet is not residing at the center of the explosion and instead is associated with the fast-moving outer ejecta. This interpretation is supported by the smooth evolution of this feature over time from near-peak to our late-phase spectra. For normal \sne, this feature has faded by $\lesssim 100~\rm{days}$ after maximum and is not usually observed in nebular-phase \sn spectra \citep[e.g., ][]{maguire2016, tucker2020}. 

With these new observations, we expand on the analysis by \citet{miller2020} and attempt to build a self-consistent explosion model for \name. The violent merger of two C/O WDs and interaction with a nearby non-degenerate companion are unlikely explanations for the early UV/optical flux excess, as our spectra place high-confidence non-detections on the expected emission from such events. Below, we discuss the remaining theories: a He-driven double-detonation (\S\ref{subsec:discuss.doubledet}), mixing of $^{56}$Ni into the outer layers of the ejecta (\S\ref{subsec:discuss.56Ni}), and the presence of a H- and He-deficient CSM from a C/O+He WD merger (\S\ref{subsec:discuss.CSM}).

\subsection{Double-Detonation Models}\label{subsec:discuss.doubledet}

Double-detonation explosions occur when a surface shell of He ignites, driving a shock wave into the WD interior and igniting the C/O core \citep{livne1990, livne1991, livne1995}. The He shell can be acquired through accretion from a nearby He star (i.e., the SD scenario; \citealp{bildsten2007}) or accretion of a tidally-disrupted lower-mass WD (i.e., the DD scenario; \citealp{fink2007}). The SD channel is unlikely due to our non-detection of H or He from the stripped material (\S\ref{subsec:spec.stripmaterial}) but the DD channel is still a possibility, especially since it can produce an early excess flux when the surface He ignites and creates $^{56}$Ni in the outer ejecta \citep{livne1995}. \citet{miller2020} fit the early light curve with double-detonations models from \citet{polin2019a} resulting in a best-fit $M_{\rm{He}} = 0.04~M_\odot$ and $M_{\rm{tot}}=0.96~M_\odot$ with the caveat that the associated model spectra predict lower photospheric velocities and more line-blanketing than are observed in the early spectra of \name.

\citet{siebert2020} derive a higher total mass of $M_{\rm{tot}}=1.15~M_\odot$ with a shell mass of $M_{\rm{He}} = 0.05~M_\odot$ by comparing the observed nebular spectrum to the nebular spectra models of \citet{polin2019}. However, this result is driven by the \CaII/\FeIII ratio (see Figs. 4 and 7 from \citealp{polin2019}) and this model both severely over-predicts the peak luminosity and fails to reproduce the UV/optical ``bump'' in the early light curve (i.e., Fig. \ref{fig:earlyLC}). \citet{siebert2020} attribute the discrepancy between the photometric and spectroscopic modeling results to viewing-angle effects, as double-detonation explosions are expected to be asymmetric and have observed properties that depend on viewing angle \citep[e.g., ][]{fink2010,kromer2010,sim2012, gronow2020}. 

Observing a thick He shell double-detonation directly along the pole as suggested by \citet{siebert2020} can reconcile the lower peak luminosity with the high \SiII velocity, as the He burning creates high-velocity He ashes above the expanding \sn ejecta, which in turn suppresses the escape of photons from the inner ejecta \citep{kromer2010}. However, this then produces the largest effects on the early photometric and spectroscopic evolution \citep[e.g., ][]{kromer2010, gronow2020} as the lower peak luminosity is driven by large amounts of line-blanketing from the He ashes \citep{kromer2010, polin2019a}. Additionally, the extra absorption from the He ashes leads to other observable consequences such as extending the rise time to maximum and decreasing $\Delta m_{15}$ \citep[e.g., Fig. 8 from][]{kromer2010} with the effects increasing for higher He shell masses. Since \name has a rise time to maximum consistent with normal \sne (cf. Fig. \ref{fig:earlyLC}), a rapid decline rate \citep{miller2020, siebert2020}, and a lack of strong line-blanketing in the early optical spectra \citep{miller2020}, a pole-on viewing angle seems unlikely to reconcile the discrepant double-detonation modeling results from \citet{miller2020} and \citet{siebert2020}.

For completeness, we also briefly consider thin He shell double-detonation models. Recent simulations have shown that very low amounts of surface He can induce a detonation \citep[e.g., ][]{shen2014} although the minimum He-shell mass needed for core detonation is still unclear \citep{glasner2018}. If very low mass He-shells can induce a core detonation, the effects on observed near-peak properties may be minor \citep[e.g., ][]{townsley2019} although this conclusion has not been extensively tested in the nebular phase. The main issue with a thin He shell double-detonation producing \name is the lack of a prominent early flux excess in these models, and thin He shell double-detonations are generally considered a potential explosion mechanism for normal \sne \citep[e.g., ][]{shen2014,townsley2019} instead of peculiar events like \name. Additionally, these models exhibit a strong correlation between peak magnitude (a proxy for the WD mass) and \SiII velocity at maximum light \citep{shen2018, polin2019a}. \name does not conform to this relation, exhibiting very high \SiII velocities \citep{miller2020} but having a peak magnitude fainter than predicted (i.e., observed $M_g\approx -18.5$~mag versus the predicted $M_B\approx -19.5$~mag from \citealp{polin2019a} for $v_{\rm{SiII}}\approx -15\,000~\kms$). Thus, both thin and thick He shell double-detonation models are unable to qualitatively reproduce the observed characteristics of \name.

An interesting avenue of speculation for double-detonation models, especially the thick He shell model preferred by \citet{siebert2020}, is the presence or absence of \ion{He}{1}$\lambda1.083~\um$. Unless the He-shell is entirely consumed during the initial surface ignition, residual He material will remain in the system and could produce observable signatures. For example, \citet{boyle2017} predict \HeINIRlong absorption near maximum light whereas \citet{dessart2015} predict strong optical and NIR \ion{He}{1} emission at $\gtrsim 50~\rm{days}$ after maximum. These studies employ very different treatments of the He mass distribution so comparisons are limited, but this highlights the uncertainty of both 1) the amount of He material remaining after surface ignition and 2) the effect of any surviving He on the optical and NIR spectral evolution. Helium is a notoriously difficult element to model as it requires a full non-LTE treatment \citep[e.g., ][]{lucy1991, hachinger2012} so we consider this a promising avenue of further study. Any double-detonation model for \name not consuming all available surface He must account for its absence in near-maximum spectra \citep{miller2020}, late-phase optical (this work, \citealp{siebert2020}), and nebular NIR (this work) spectra. 

Finally, the presence of strong \CaII is not a direct consequence of a double-detonation as suggested by \citet{miller2020} (and subsequently by \citealp{siebert2020}), but instead an indicator that the Ca-rich material is located in the same region of the ejecta as the $^{56}$Ni decay products \citep[e.g., ][]{wilk2020}. \CaII is an extremely effective coolant due to its large oscillator strength, and therefore \CaII emission can be generated by any explosion model producing Ca$^+$ in the ejecta regardless of the explosion mechanism itself. Several models in the literature produce strong \CaII emission in the nebular phase without requiring a double-detonation explosion \citep[e.g., ][]{mazzali2012, blondin2017, botyanszki2017, galbany2019, wilk2020}. These conditions \textit{can} be created by a double-detonation explosion \citep[e.g., ][]{polin2019}, as the He burning naturally intersperses \fsNi with intermediate-mass elements such as Ca, but this effect is not exclusive to double-detonation explosions. Instead, the detection of Ca provides insight into the chemical distribution and bulk ionization state of the ejecta.

\subsection{Ni Mixing Models}\label{subsec:discuss.56Ni}

Next, we consider $^{56}$Ni mixing as a potential explanation for both the early flux excess and the presence of \CaII in the nebular spectra. The source of the early flux excess is the same for both $^{56}$Ni-mixing and double-detonation models: the presence of $^{56}$Ni in the outer ejecta provides excess heating and thus excess flux \citep{piro2012, piro2016}. However, the outer Ni in double-detonation models is produced by the He burning process whereas $^{56}$Ni-mixing requires a transport mechanism to move the inner $^{56}$Ni material to the outer ejecta. Previously-invoked mechanisms include irregular/asymmetric deflagrations \citep{khokhlov1993, hoeflich1996}, gravitationally-confined detonations (GCDs, \citealp{plewa2004, piro2012}), or the direct collision of two WDs \citep{rosswog2009, raskin2009}. Each theory has observational inconsistencies with normal \sne but these problems may not apply to peculiar \sne such as \name.

It is plausible for \name that the outer $^{56}$Ni clumps responsible for the early flux excess can also account for the detection of \CaII in the nebular phase. Outward mixing of $^{56}$Ni may boost the \CaII emission due to the energy from local positron deposition in the Ca-rich region and ensuing radiative cooling. However, there are no published models that address this scenario and how $^{56}$Ni mixing affects the nebular spectra. The presence of \CaII emission in the nebular spectra models of \citet{wilk2020} is highly dependent on the amount of clumping in the ejecta, providing a potential avenue for $^{56}$Ni mixing to address both the early early flux excess and the nebular \CaII emission. Furthermore, the $^{56}$Ni mixing models of \citet{kasen2006} predict no secondary $I$-band maximum for a fully-mixed composition, a low $^{56}$Ni yield, or a combination of these two. This could explain the extremely weak NIR secondary maximum in \name seen in Fig. \ref{fig:TESS}. \citet{miller2020} estimate a low $^{56}$Ni mass of $0.31\pm0.05~M_\odot$ which is lower than for normal \sne ($0.4-0.8~M_\odot$, \citealp{scalzo2014}). Thus, $^{56}$Ni mixing is a promising avenue for reproducing some of the major aspects of \name and studies of normal \sne have produced promising results for the existence of shallow \fsNi \citep[e.g., ][]{piro2014}. 

However, there is a complex interplay between the necessary synthesized $^{56}$Ni to power the optical light curve \citep[e.g., ][]{arnett1982}, the amount of mixing needed to reproduce the early flux excess \citep[e.g., ][]{piro2016, magee2020}, the amount of mixing needed to suppress the secondary NIR maximum \citep[e.g., ][]{kasen2006}, and limitations on outer $^{56}$Ni from the lack of unusual spectroscopic features (i.e., line-blanketing effects) in the early spectroscopic evolution \citep{miller2020}. These features are not self-consistently addressed by any model in the literature nor are there predictions for the effect on nebular spectra. New models are needed to determine if the right combination of $^{56}$Ni mixing can produce all aspects of \name. Additionally, $^{56}$Ni mixing is a result of the WD explosion, not a cause, and an explosion mechanism must still be invoked to actually destabilize the WD. All explosion models struggle to reconcile the low luminosity of \name with the high \SiII velocity, a hindrance for any model invoking \fsNi to explain the peculiarities of \name.

\subsection{H- and He-Deficient CSM}\label{subsec:discuss.CSM}

A third possibility for excess flux soon after explosion is the presence of a dense CSM surrounding the exploding WD. The near-peak spectra from \citet{miller2020} and our nebular spectra exclude any H or He emission at high confidence, so any CSM must be H and He depleted. This can be created by a WD merger where the more massive WD tidally disrupts and accretes the lower-mass WD \citep{fink2007, pakmor2010}. This process is not completely efficient and some material will escape into the surrounding ISM instead of being consumed by the explosion \citep[e.g., ][]{shen2013}. H- and He-rich rich surface layers comprise only a small fraction of the total C/O WD mass ($M_{\rm{H}} / M_{\rm{WD}} \lesssim 10^{-4}$ and $M_{\rm{He}}/M_{\rm{WD}} \sim 10^{-2}$; e.g., \citealp{romero2012}) so inefficient accretion/mass transfer can readily produce a H- and He-deficient CSM.

\citet{miller2020} showed that the $0.9+0.76~M_\odot$ merger model from \citet{kromer2016} can qualitatively reproduce the photometric evolution of \name, although the early flux excess itself was not modeled due to the numerous possible CSM configurations. The ejecta-CSM interaction models from \citet{piro2016} can produce an early flux excess with duration $< 4$~days, blue colors, and peak $V$-band luminosity of $M_V\sim -15$~mag for a CSM radius of $\sim 10^{12}~\rm{cm}$, similar to the early flux excess observed in \name. However, the primary issue with a violent merger of two C/O WDs producing \name is the lack of prominent O emission lines in our optical and NIR spectra. While ignition of the primary WD produces \fsNi and iron-group elements, the secondary lower-mass WD is only partially burnt \citep{pakmor2011}. For a C/O secondary WD, this results in unburnt oxygen near the center of the ejecta which produces strong O emission lines in the nebular spectra as seen in SN~2010lp \citep{taubenberger2013, kromer2013}. 

Interestingly, \citet{kromer2013} and \citet{taubenberger2013} state that [\ion{O}{1}] is not expected for a merger between a C/O WD and a He WD, as there is no unburnt oxygen material to produce the nebular emission lines. There are no models directly assessing the result of a C/O WD merging with a He WD, but we consider two simplistic outcomes: the He WD is partially burnt or the He WD experiences no burning. If the He WD is partially burnt, by-products such as O, Ca, and Ti should be produced and reside near the center of the ejecta. This matches the detection of \CaIIlong, but the burning process must be highly efficient otherwise residual O or He remains in the ejecta and should produce nebular emission lines. If the He WD experiences little or zero burning, a large amount of unburnt He material remains near the center of ejecta. This is reminiscent of material stripped from a He donor star and should produce strong NIR He emission lines \citep[e.g., ][]{botyanszki2018} which we exclude at high confidence.

Thus, a H- and He-deficient CSM produced in the merger of two WDs is an unlikely progenitor system to explain \name. A merger of two C/O WDs should produce O emission which we do not detect, and a merger of a C/O WD with a He WD would likely produce strong emission lines of O and/or He. The exception is if the He WD experiences significant burning and converts all He to elements heavier than O which would produce nebular \CaII emission but no O or He emission lines. We consider this scenario unlikely but numerical simulations should confirms these qualitative considerations. 

Finally, there is the direct collision of two WDs \citep[e.g., ][]{rosswog2009, raskin2009}. This scenario is thought to produce double-peaked emission lines of \fsNi decay products \citep[e.g., ][]{dong2015} as the ensuing velocity distribution is inherently bimodal. However, double-peaked or asymmetric emission lines can also be produced by off-center explosions \citep[e.g., ][]{gall2018, vallely2019}. We do not find any evidence for double-peaked or flat-topped emission profiles in our optical spectra of \name, but our optical spectra are not late enough in the nebular phase for a definitive conclusion. Direct collision models in the literature are scarce and important questions, such as the presence of unburnt material near the center of the ejecta, are currently unanswered. If unburnt material survives the collision and subsequent explosion, we would expect similar spectral signatures as the merger scenario (i.e., He or O emission lines). We encourage further modeling of this scenario to understand the chemical composition of the surviving material and its effect on the nebular spectra.

\section{Conclusion}\label{sec:conclusion}

We have presented and analyzed new photometric and spectroscopic observations of the unusual \sn 2019yvq which exhibited a early blue flux excess, low peak luminosity, high \SiII velocities, and the presence of \CaIIlong and \CaIINIR NIR triplet emission in the nebular spectra. Our near-explosion $g$-band photometry places new constraints on the duration of the early flux excess and the time of first light. The \textit{TESS} light curve reveals a weak secondary maximum, atypical for intermediate-luminosity \sne. The NIR spectrum excludes any stripped material from a nearby non-degenerate companion due to the lack of H and He emission lines. The optical spectra confirm these non-detections and reveal prominent \CaII and \CaIINIR NIR triplet emission. The \CaIIlong line declines in absolute flux between our two epochs but strengthens relative to the surrounding \NiII and \FeII lines whereas the \CaIINIR NIR triplet shows little change between the two epochs.

No explosion model in the literature addresses all of the aspects of \name. One commonality between the models is the difficulty reconciling the high photospheric velocities with the low peak luminosity. The early flux excess can be explained with several physical scenarios (see \citealp{miller2020} for an in-depth discussion) with our new observations strongly disfavoring interaction with a non-degenerate companion or a violent merger of two C/O WDs. However, the other potential sources for the flux excess (a double-detonation explosion, \fsNi-mixing into the outer ejecta, and a H-/He-deficient CSM) also have shortcomings or contradictions (\S\ref{sec:discuss}).

\name highlights the uncertainties plaguing \sn explosion models. Even with the exquisite photometric and spectroscopic observations from \citet{miller2020}, \citet{siebert2020}, and this work, \name remains an enigma. However, these high-quality observations provide an excellent foundation for future modeling attempts as \name enters the growing taxonomy of WD explosions.

\vspace{1cm}
\section*{Acknowledgements}

We thank Connor Auge and Jason Hinkle for providing useful comments on the manuscript. 
 
M.A.T. acknowledges support from the DOE CSGF through grant DE-SC0019323. G.S.A. acknowledges support from an award from the Space Telescope Science Institute in support of program SNAP-15922. C.A. and B.J.S. are supported by NASA grant 80NSSC19K1717 and NSF grants AST-1920392 and AST-1911074.  B.J.S., and C.S.K. are supported by NSF grant AST-1907570. C.S.K. is supported by NSF grant AST-181440.

We thank the Las Cumbres Observatory and its staff for its continuing support of the ASAS-SN project. ASAS-SN is supported by the Gordon and Betty Moore Foundation through grant GBMF5490 to the Ohio State University, and NSF grants AST-1515927 and AST-1908570. Development of ASAS-SN has been supported by NSF grant AST-0908816, the Mt. Cuba Astronomical Foundation, the Center for Cosmology and AstroParticle Physics at the Ohio State University, the Chinese Academy of Sciences South America Center for Astronomy (CAS- SACA), and the Villum Foundation. 

This work was enabled by observations made from the Gemini North telescope, located within the Maunakea Science Reserve and adjacent to the summit of Maunakea. We are grateful for the privilege of observing the Universe from a place that is unique in both its astronomical quality and its cultural significance.

Based on observations obtained at the international Gemini Observatory, a program of NSF’s NOIRLab, which is managed by the Association of Universities for Research in Astronomy (AURA) under a cooperative agreement with the National Science Foundation. on behalf of the Gemini Observatory partnership: the National Science Foundation (United States), National Research Council (Canada), Agencia Nacional de Investigaci\'{o}n y Desarrollo (Chile), Ministerio de Ciencia, Tecnolog\'{i}a e Innovaci\'{o}n (Argentina), Minist\'{e}rio da Ci\^{e}ncia, Tecnologia, Inova\c{c}\~{o}es e Comunica\c{c}\~{o}es (Brazil), and Korea Astronomy and Space Science Institute (Republic of Korea).

Some of he data presented herein were obtained at the W. M. Keck Observatory, which is operated as a scientific partnership among the California Institute of Technology, the University of California and the National Aeronautics and Space Administration. The Observatory was made possible by the generous financial support of the W. M. Keck Foundation.

\bibliography{sample63}{}
\bibliographystyle{aasjournal}



\end{document}

%% file: obs-info.tex
\begin{table*}
    \centering
    \caption{Details of the new spectra observations.}
    \begin{tabular}{cccccccc}
        Instrument & Telescope & MJD & Phase [d] & Exp. time [s] & Range & $\lambda/\Delta\lambda$ & Airmass \\\hline
        GMOS-N & Gemini-North & 58991.382 & +128 & 3600 & $4500-9100~\rm\AA$ & $\sim2000$ & 1.60 \\
        GMOS-N & Gemini-North & 59013.279 & +150 & 4800 & $4500-9100~\rm\AA$ & $\sim2000$ & 1.48 \\
        NIRES & Keck II & 59038.263 & +173 & 2400 & $0.95-2.5~\rm\mu \rm m$ & $\sim 2700$ & 1.63 \\
        \hline
    \end{tabular}
    \label{tab:obsinfo}
\end{table*}

%% file: newphot.tex
\begin{table}
    \centering
    \caption{New ASAS-SN and TESS photometry of \name. A portion of the data is provided to illustrate the format and content. The full light curve is included with the online version of the manuscript.}
    \begin{tabular}{lrr}
        Source & MJD & Flux [mJy] \\\hline
        ASAS-SN & 58812.47524 & $0.04\pm0.04$ \\
        ASAS-SN & 58823.48662 & $0.03\pm0.03$ \\
        ASAS-SN & 58824.46758 & $0.06\pm0.04$ \\
        ASAS-SN & 58842.39607 & $0.10\pm0.04$ \\
        $\ldots$ & $\ldots$ & $\ldots$ \\\hline
    \end{tabular}
    \label{tab:newphot}
\end{table}